# A DISTRIBUTED ALGORITHM FOR PERSONAL SOUND ZONES SYSTEMS

*Sipei Zhao[1], Guoqiang Zhang[1], Eva Cheng[1] and Ian S. Burnett[2]*

[1]Centre for Audio, Acoustics and Vibration, Faculty of Engineering and IT, University of Technology Sydney, Ultimo, NSW 2007, Australia
[2]DVC and VP STEM, RMIT University, Melbourne, VIC 3000, Australia

Sipei.Zhao@uts.edu.au, Guoqiang.Zhang@uts.edu.au, Eva.Cheng@uts.edu.au, Ian.Burnett2@rmit.edu.au

## ABSTRACT

A Personal Sound Zones (PSZ) system aims to generate two or more independent listening zones that allow multiple users to listen to different music/audio content in a shared space without the need for wearing headphones. Most existing studies assume that the acoustic paths between loudspeakers and microphones are measured beforehand in a stationary environment. Recently, adaptive PSZ systems have been explored to adapt the system in a time-varying acoustic environment. However, because a PSZ system usually requires multiple loudspeakers, the multichannel adaptive algorithms impose a high computational load on the processor. To overcome that problem, this paper proposes an efficient distributed algorithm for PSZ systems, which not only spreads the computational burden over multiple nodes but also reduces the overall computational complexity, at the expense of a slight decrease in performance. Simulation results with true room impulse responses measured in a Hemi-Anechoic chamber are performed to verify the proposed distributed PSZ system.

## 1. INTRODUCTION

Due to promising applications in museums, home theatres, vehicle cabins, mobile audio devices, and outdoor musical events *etc*., Personal Sound Zones (PSZ) systems have been broadly studied in the past twenty years[1]. Popular methods for PSZ systems include the Acoustic Contrast Control (ACC) [2], the Pressure Matching (PM) [3], and Mode Matching (MM) [4] *etc*.

The ACC method generates a high acoustic isolation between zones, but fails to produce a uniform sound field in the bright zone, thereby resulting in poorer sound quality [5]. On the other hand, the PM and MM methods reproduce a target sound field, such as a plane wave, in the bright zone, thus improving the sound quality. However, the two methods usually generate a lower acoustic contrast between the bright and dark zones compared to the ACC method. To combine the advantages of the ACC and PM methods, a weighted Pressure Matching (wPM) technique has been studied in [6], [7]. It has also been shown that the ACC and PM methods are two special cases under the framework of the variable span linear filters [8].

The abovementioned studies are based on the assumption that the PSZ system works in a stationary acoustic environment and the acoustic plants are known or measured in advance. Recently, adaptive PSZ systems have been explored to adapt the PSZ system to changes in environments [9]–[12]. Vindrola *et al*. [9] uses the Filtered-reference Least Mean Square algorithm (F*x*LMS) to adapt the PSZ system. Zhao and Burnett [10] proposed an adaptive wPM-based PSZ system with online plant modelling to track variations in acoustic paths. Similarly, Hu *et al*. [11] proposed to maintain a fixed acoustic contrast with online tracking of acoustic transfer functions. Gao *et al*. [12] explored multizone sound field reproduction with adaptive control of scattering effects of listeners in the sound zones. These studies take a centralized algorithm running on a single processor. However, in a PSZ system, usually multiple loudspeakers are needed to produce high acoustic isolation between zones. Therefore, the computational burden would be high for a single processor.

To overcome this limitation, distributed PSZ systems will be investigated in this paper. In the past decade, distributed processing over acoustic sensor and actuator networks have been widely explored for active noise control, either non-cooperative [13], [14] or cooperative based on incremental [15], [16] and diffusion [17], [18] combinations. Van Rompaey and Moonen investigated distributed ACC using gradient based Generalized Eigenvalue Decomposition (GEVD) [19] and distributed PM based on the Gauss-Seidel algorithm [20] for PSZ systems.

In an alternative approach, this paper proposes an efficient distributed wPM algorithm for PSZ systems. A distributed implementation of the centralized wPM algorithm is presented first, and then an efficient distributed algorithm is proposed to reduce both overall computation and communication load over the entire network. Simulations with measured room impulse responses are carried out to verify the proposed algorithm.

## 2. CENTRALIZED PSZ

A centralized PSZ system is illustrated in Fig. 1, where an array of $L$ loudspeakers generates sound signals that are monitored with multiple microphones in both the bright and dark zones. The microphone signals are fed back to a central processor to update the control filters of all the loudspeakers to track any changes in acoustic environments. The objective of the central processor is to optimize the Finite Impulse Response (FIR) filters $\mathbf{w}_l$ (for $l$ from 1 to $L$) to produce a high acoustic energy in the bright zone while suppressing the sound pressure levels in the dark zone. A sound signal to be reproduced in the bright zone, $x[n]$ ($n$ is the time instant), goes through the control filters before sending to the loudspeakers. Hence the signal fed to the $l$-th loudspeaker is [10]

$$u_l(n) = \mathbf{x}(n)^T \mathbf{w}_l, \qquad (1)$$

where $\mathbf{x}(n) = [x(n), x(n-1), \ldots, x(n-K+1)]^T$ and $\mathbf{w}_l = [w_l(1), w_l(2), \ldots, w_l(K)]^T$, with $K$ being the length of the control filters.

The output signals of the loudspeakers will propagate through the acoustic plants before arriving at the microphones in either the bright or dark zones. Without loss of generality, the FIR filters $\mathbf{h}_{B,ml}$ and $\mathbf{h}_{D,ml}$, both with a length of $J$, are employed to model the plants from the $l$-th loudspeaker to the $m$-th microphone in the bright and dark zones, respectively. Therefore, the sound pressure at the $m$-th microphone in the bright zone is [10]

$$p_{B,m}(n) = \sum_{l=1}^{L} \mathbf{h}_{B,ml}^T \mathbf{u}_l(n), \qquad (2)$$

where $\mathbf{h}_{B,ml} = [h_{B,ml}(0), h_{B,ml}(1), \ldots, h_{B,ml}(J-1)]^T$ and $h_{ml}(j)$ is the $j$-th coefficient of the acoustic path from the $l$-th loudspeaker to the $m$-th microphone, and $\mathbf{u}_l(n) = [u_l(n), u_l(n-1), \ldots, u_l(n-J+1)]^T$.

By substituting Eq. (1) into Eq. (2), we obtain

$$p_{B,m}(n) = \sum_{l=1}^{L} \mathbf{h}_{B,ml}^T \mathbf{X}(n) \mathbf{w}_l, \qquad (3)$$

where $\mathbf{X}(n) = [\mathbf{x}(n), \mathbf{x}(n-1), \ldots, \mathbf{x}(n-J+1)]^T$. Stacking the sound pressure measured by all the microphones in the bright zone yields

$$\mathbf{p}_B[n] = \begin{bmatrix} \mathbf{r}_{B,11} & \cdots & \mathbf{r}_{B,1L} \\ \vdots & \ddots & \vdots \\ \mathbf{r}_{B,M1} & \cdots & \mathbf{r}_{B,ML} \end{bmatrix} \begin{bmatrix} \mathbf{w}_1 \\ \vdots \\ \mathbf{w}_L \end{bmatrix} = \mathbf{R}_B \mathbf{w}, \qquad (4)$$

where $\mathbf{r}_{B,ml} = \mathbf{h}_{B,ml}^T \mathbf{X}(n)$ are the filtered input signals based on the assumption that the control filters change slowly compared to the acoustic paths. The sound pressure in the dark zone can be similarly expressed, as shown in [10]. The cost function for wPM-based PSZ system is defined as [10]

$$J(\mathbf{w}) = \mathbb{E}\left\{ \kappa (\mathbf{p}_B - \mathbf{p}_T)^T (\mathbf{p}_B - \mathbf{p}_T) + (1-\kappa) \mathbf{p}_D^T \mathbf{p}_D \right\}, \qquad (5)$$

where $\mathbf{p}_T$ is the target sound in the bright zone and $0 < \kappa < 1$ is a weighting factor.

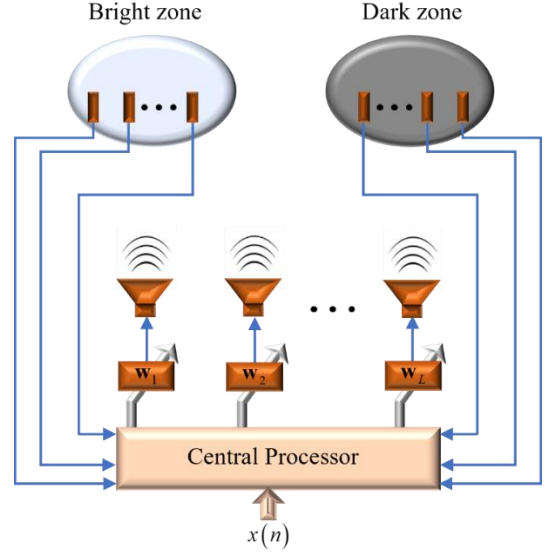

**Fig. 1**. Block diagram of a centralized PSZ system.

By following the least mean square procedure, the update equation for the $l$-th control filter is derived as

$$\begin{aligned} \mathbf{w}_l(n+1) &= \mathbf{w}_l(n) \\ &- \mu \sum_{m=1}^{M} \left[ \kappa \tilde{\mathbf{r}}_{B,ml}^T (p_{B,m} - p_{T,m}) + (1-\kappa) \tilde{\mathbf{r}}_{D,ml}^T p_{D,m} \right], \end{aligned} \qquad (6)$$

where $\mu$ is the step size and $\tilde{\mathbf{r}}_{B,ml}$ and $\tilde{\mathbf{r}}_{D,ml}$ are the filtered input signals through the estimated plant model from the $l$-th loudspeaker to the $m$-th microphone in the bright and dark zones, respectively. The plant models can be estimated using either offline [17] or online [10] methods. For simplicity, this paper assumes offline estimation of the acoustic plants is performed in a modelling stage before real-time control.

It is noted that, in the adaptive algorithm in (6), updating the control filter for each loudspeaker requires measured sound pressures from all the microphones, i.e., $p_{B,m}$ and $p_{D,m}$ for $m$ from 1 to $M$. Therefore, a centralized processor is needed to gather the information from all the microphones and adapt the control filters. To distribute the computational burden over multiple node processors, a distributed algorithm is formulated in the next section.

## 3. DISTRIBUTED PSZ

A distributed PSZ system (Fig. 2) employs multiple node processors (numbered from 1 to $L$) to replace the single central processor. Each node processor can calculate the filter coefficients based on its own microphone signals and communicate with other node processors to share information. For simplicity, in the following formulation, it is assumed that the number of microphones in each zone is equivalent to that of loudspeakers (i.e., $M = L$), but the algorithm can be extended to include multiple loudspeakers and/or microphones in each node.

The update equations of all the control filters are stacked in a vector as

$$\mathbf{w}(n+1) = \mathbf{w}(n) - \mu \sum_{m=1}^{M}\left[\kappa \tilde{\mathbf{R}}_{B,m}\left(p_{B,m} - p_{T,m}\right) + (1-\kappa)\tilde{\mathbf{R}}_{D,m} p_{D,m}\right], \quad (7)$$

where $\tilde{\mathbf{R}}_{B,m} = \begin{bmatrix} \tilde{\mathbf{r}}_{B,m1} & \tilde{\mathbf{r}}_{B,m2} & \cdots & \tilde{\mathbf{r}}_{B,mL} \end{bmatrix}^{T}$ and $\tilde{\mathbf{R}}_{D,m}$ is similarly defined.

To share information between nodes, two strategies i.e., incremental [21] and diffusion [22], are usually used. This paper formulates the distributed personal audio system based on the diffusion strategy due to its robustness to link failure and capacity for parallel computing. To formulate the adaptive algorithm over a distributed network, (7) is rearranged by multiplying $M$ to both sides and introducing $\mu_1 = \mu M$, i.e.,

$$\mathbf{w}(n+1) = \frac{1}{M}\sum_{m=1}^{M}\left(\mathbf{w}(n) - \mu_1 \mathbf{v}_m\right), \quad (8)$$

where $\mathbf{v}_m(n) = \kappa \tilde{\mathbf{R}}_{B,m}\left(p_{B,m} - p_{T,m}\right) + (1-\kappa)\tilde{\mathbf{R}}_{D,m} p_{D,m}$. The update equation (8) can be divided into two steps following the Adapt-then-Combine (ATC) scheme, i.e., the adaptation step,

$$\boldsymbol{\varphi}_m(n) = \mathbf{w}(n) - \mu_1 \mathbf{v}_m \text{ for } m = 1, 2, \ldots, M, \quad (9)$$

and the combination step,

$$\mathbf{w}(n+1) = \sum_{m=1}^{M}\frac{1}{M}\boldsymbol{\varphi}_m(n). \quad (10)$$

It is clear that in the adaptation step (9), each node processor estimates the control filters based on its own microphone signals, making it feasible to distribute the overall computational burden over multiple node processors. Meanwhile, in the combination step (10), the node processors share the estimations with each other for data fusion before updating the global control filters. The algorithmic behavior of (9) and (10) are the same as the centralized version; therefore, they are referred to as the distributed implementation of the centralized algorithm. It is worth noting that the combination in (10) requires the estimations for all control filters by all nodes, which applies a high communication overhead to the network in addition to the high computational burden of each node.

To overcome this limitation, an efficient distributed PSZ system is proposed. Instead of estimating all the control filters, each node processor only estimates the control filters of its neighbor nodes, including itself. To do so, the adaptation step (9) is modified as

$$\hat{\boldsymbol{\varphi}}_m(n) = \hat{\mathbf{w}}_m(n) - \mu\, \hat{\mathbf{v}}_m, \quad (11)$$

where $\hat{\mathbf{w}}_m(n) = \text{col}\{\mathbf{w}_l(n)\}, l \in \mathcal{N}_m$ is an augmented vector that collocates the control filters within the neighborhood of the $m$-th node, $\mathcal{N}_m$, and

$$\hat{\mathbf{v}}_m = \kappa \hat{\mathbf{R}}_{B,m}\left(p_{B,m} - p_{T,m}\right) + (1-\kappa)\hat{\mathbf{R}}_{D,m} p_{D,m}, \quad (12)$$

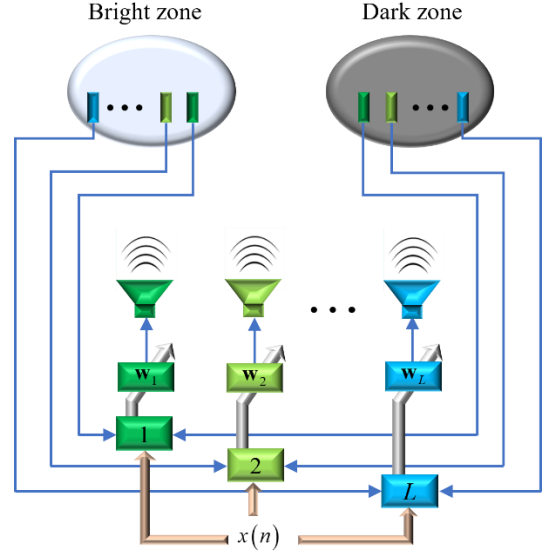

Fig. 2. Block diagrams of a distributed PSZ system.

where $\hat{\mathbf{R}}_{B,m} = \text{col}\{\tilde{\mathbf{r}}_{B,ml}^{T}\}, l \in \mathcal{N}_m$ and $\hat{\mathbf{R}}_{D,m}$ is similarly defined.

It is noted that the adaptation outcome of each node $\hat{\boldsymbol{\varphi}}_m(n) = \text{col}\{\tilde{\mathbf{w}}_m^l(n)\}, l \in \mathcal{N}_m$ is also an augmented vector consisting of estimations of its neighbor nodes, where $\tilde{\mathbf{w}}_m^l(n)$ denotes the $l$-th node's control filter estimated using the $m$-th node's microphone signals. In the modified combination step, different estimations of the $l$-th node's control filter by its neighbor nodes are shared and fused, i.e.,

$$\hat{\mathbf{w}}_l(n+1) = \sum_{m \in \mathcal{N}_l} \alpha_{ml} \tilde{\mathbf{w}}_m^l(n), \quad (13)$$

where $\alpha_{ml}$ is the combination coefficients.

The modified adaptation in (11) and combination in (13) form a complete cycle of the efficient distributed PSZ algorithm, which not only reduces the computation complexity of each node but also decreases the communication overhead of the network. By comparing the adaptation steps (9) and (11), the number of multiplications for each node is reduced from $2[(J+1)KL + 1]$ for the distributed implementation of the centralized algorithm to $2[(J+1)K|\mathcal{N}_m|+1]$ for the proposed efficient algorithm. Since $JK$ is much larger than 1 and $|\mathcal{N}_m|$ is smaller than $L$, the proposed efficient distributed algorithm reduces each node's computational complexity by $|\mathcal{N}_m|/L$ times. On the other hand, by comparing the combination steps in (10) and (13) the communication overhead over the entire network is reduced significantly from $KL^2(L-1)$ to $K\sum_{m=1}^{M}(|\mathcal{N}_m|-1)$.

## 4. SIMULATIONS AND DISCUSSIONS

To verify the proposed distributed personal audio system, simulations with true Room Impulse Responses (RIRs) measured in the UTS Hemi-Anechoic Chamber were performed. The experimental setup is shown in Fig. 3(a), where the RIRs from 60 loudspeakers on a circular truss with a diameter of 1.5 m to 64 microphones ($8 \times 8$ array, with a microphone interval of 40 mm) in each zone were measured. Detailed description of the experimental setup and measurement procedure is available in [23].

For the simulations in this paper, to reduce calculation load, only eight loudspeakers (those in the yellow rectangles in Fig. 3(a)) and eight microphones in each zone (the first eight microphones in the planar array described in [23]) were used. As illustrated in Fig. 3(b), each node consisted of one loudspeaker, one microphone in the bright zone, and one microphone in the dark zone. The adjacent nodes can share information with each other, as shown by the blue lines in Fig. 3(b). Without loss of generality, all the combination coefficients for all the node links were set to 1.

In the simulations, all the acoustic paths were modeled as a 128-tap Finite Impulse Response (FIR) filter at a sampling rate of 4 kHz. The length of the control filters was also 128 taps. A bandpass filtered white Gaussian noise in the frequency range of 100 Hz and 1,000 Hz was used as the input signal and the step sizes for the centralized and the proposed distributed algorithm were 0.06 and 0.016, respectively. The target signal in the bright zone is a delayed version of the signal from the first loudspeaker. The performance of the PSZ system is evaluated by the Mean Square Error (MSE) in the bright zone and the Acoustic Contrast (AC) between the two zones, the definitions of which are available in [10].

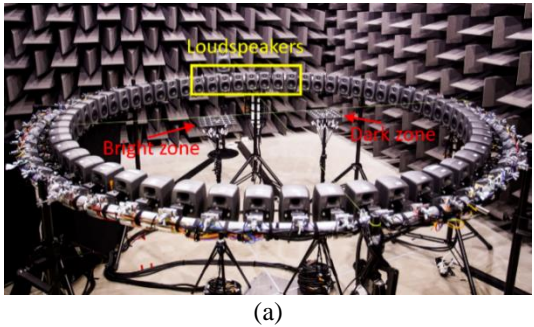

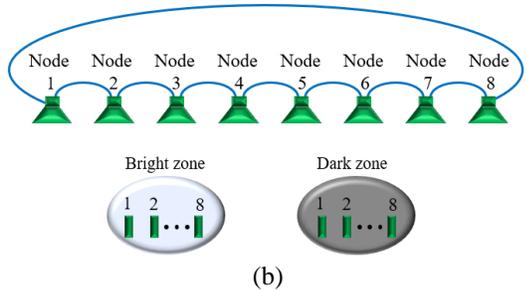

**Fig. 3**. (a) Photo of the experimental setup and (b) network topology for the simulations.

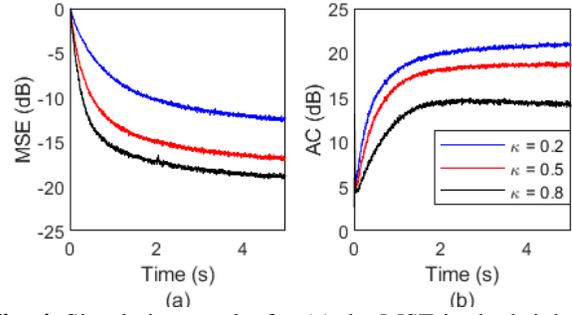

**Fig. 4**. Simulation results for (a) the MSE in the bright zone and (b) the AC between the bright and dark zones.

The simulation results for the proposed algorithm are shown in Fig. 4 for different weight factor $\kappa$. It is clear from Fig. 4(a) that, with the increase in the weighting factor $\kappa$, the learning curve for MSE converges faster and the steady-state MSE decreases. The steady-state MSEs are $-12.4$ dB, $-16.9$ dB and $-19.2$ dB for $\kappa = 0.2$, $\kappa = 0.5$ and $\kappa = 0.8$, respectively. On the other hand, as shown in Fig. 4(b), increasing the weighting factor $\kappa$ leads to a lower-converging learning curve for the AC and a degradation in the AC. The steady-state AC for $\kappa = 0.2$, $\kappa = 0.5$ and $\kappa = 0.8$ are 20.8 dB, 18.5 dB and 14.0 dB, respectively. The weighting factor balances the sound quality in the bright zone and sound isolation between the two zones, as expected.

The above results demonstrate the efficacy of the proposed distributed PSZ system. For the current simulation setup, $|\mathcal{N}_m| = 3$ and $L = 8$, so the computational complexity of the proposed algorithm is only 37.5% of that of the centralized algorithm, and the communication overhead over the entire network is only 3.6% of that of the centralized algorithm. However, it is noted that the reduction in the computational complexity and communication cost comes at an expense of slightly decreased performance. Simulation results (not presented here due to page limit) show that the MSE of the proposed algorithm is approximately 2.5 dB lower and the AC approximately 2.0 dB higher than the centralized algorithm when $\kappa = 0.5$. More studies are needed to investigate the convergence condition, as well as the effect of the network topology and combination coefficients on the convergence behavior and steady-state performance. This will be a topic of future work.

## 5. CONCLUSIONS

This paper proposed an efficient algorithm for adaptive PSZ systems to reduce the overall computational complexity and distribute the computational burden over multiple node processors. The cost for the improvements is a slight degradation in performance. Simulation results with measured RIRs demonstrate the efficacy of the proposed algorithm although further work is needed to comprehensively understand the convergence behavior and the steady-state performance.